\documentclass[%
 aip,
 amsmath,amssymb,
 reprint,%
]{revtex4-1}
\usepackage{textcomp}
\usepackage{graphicx}
\usepackage{dcolumn}
\usepackage{bm}

\usepackage[utf8]{inputenc}
\usepackage[T1]{fontenc}
\usepackage{mathptmx}
\usepackage{siunitx}
\DeclareSIUnit{\fluence}{\milli\joule\per\centi\meter\squared}

\begin{document}

\preprint{AIP/123-QED}

\title[]{All-optical switching on the nanometer scale excited and probed with femtosecond extreme ultraviolet pulses}

\author{Kelvin Yao}
 \affiliation{Max-Born-Institut für Nichtlineare Optik und Kurzzeitspektroskopie, Max-Born-Straße 2A, 12489 Berlin, Germany}
 
\author{Felix Steinbach}
 \affiliation{Max-Born-Institut für Nichtlineare Optik und Kurzzeitspektroskopie, Max-Born-Straße 2A, 12489 Berlin, Germany}
 
\author{Martin Borchert}
 \affiliation{Max-Born-Institut für Nichtlineare Optik und Kurzzeitspektroskopie, Max-Born-Straße 2A, 12489 Berlin, Germany}
 
\author{Daniel Schick}
 \affiliation{Max-Born-Institut für Nichtlineare Optik und Kurzzeitspektroskopie, Max-Born-Straße 2A, 12489 Berlin, Germany}

\author{Dieter Engel}
 \affiliation{Max-Born-Institut für Nichtlineare Optik und Kurzzeitspektroskopie, Max-Born-Straße 2A, 12489 Berlin, Germany}
 
 \author{Filippo Bencivenga}
 \affiliation{Elettra Sincrotrone Trieste S.C.p.A., Strada Statale 14, km 163.5, 34149 Basovizza (TS), Italy}
 
\author{Riccardo Mincigrucci}
 \affiliation{Elettra Sincrotrone Trieste S.C.p.A., Strada Statale 14, km 163.5, 34149 Basovizza (TS), Italy}
 
\author{Laura Foglia}
 \affiliation{Elettra Sincrotrone Trieste S.C.p.A., Strada Statale 14, km 163.5, 34149 Basovizza (TS), Italy}
 
 \author{Emanuele Pedersoli}
 \affiliation{Elettra Sincrotrone Trieste S.C.p.A., Strada Statale 14, km 163.5, 34149 Basovizza (TS), Italy}

\author{Dario De Angelis}
 \affiliation{Elettra Sincrotrone Trieste S.C.p.A., Strada Statale 14, km 163.5, 34149 Basovizza (TS), Italy}
 
 \author{Matteo Pancaldi}
 \affiliation{Elettra Sincrotrone Trieste S.C.p.A., Strada Statale 14, km 163.5, 34149 Basovizza (TS), Italy}
 
  \author{Björn Wehinger}
 \affiliation{ESRF - The European Synchrotron, 71, Avenue des Martyrs, 38043 Grenoble, France}
 \affiliation{Department of Molecular Sciences and Nanosystems, Ca’ Foscari  University of Venice, via Torino 155, 30172 Venezia Mestre, Italy}
 
   \author{Flavio Capotondi}
 \affiliation{Elettra Sincrotrone Trieste S.C.p.A., Strada Statale 14, km 163.5, 34149 Basovizza (TS), Italy}
 
    \author{Claudio Masciovecchio}
 \affiliation{Elettra Sincrotrone Trieste S.C.p.A., Strada Statale 14, km 163.5, 34149 Basovizza (TS), Italy}
 
 \author{Stefan Eisebitt}
 \affiliation{Max-Born-Institut für Nichtlineare Optik und Kurzzeitspektroskopie, Max-Born-Straße 2A, 12489 Berlin, Germany}
 
\author{Clemens von Korff Schmising}%
 \affiliation{Max-Born-Institut für Nichtlineare Optik und Kurzzeitspektroskopie, Max-Born-Straße 2A, 12489 Berlin, Germany}
 \email{Clemens.KorffSchmising@mbi-berlin.de}

\date{\today}

\begin{abstract}
Ultrafast control of magnetization on the nanometer length scale, in particular all-optical switching, is key to putting ultrafast magnetism on the path towards future technological application in data storage technology. 
However, magnetization manipulation with light on this length scale is challenging due to the wavelength limitations of optical radiation.
Here, we excite transient magnetic gratings in a GdFe alloy with a periodicity of 87\,nm by interference of two coherent femtosecond light pulses in the extreme ultraviolet spectral range.
The subsequent ultrafast evolution of the magnetization pattern is probed by diffraction of a third, time-delayed pulse tuned to the Gd \textit{N}-edge at a wavelength of 8.3\,nm.
By examining the simultaneously recorded first and second diffraction orders and by performing reference real-space measurements with a wide-field magneto-optical microscope with femtosecond time resolution, we can conclusively demonstrate the ultrafast emergence of all-optical switching on the nanometer length scale.

\end{abstract}

\maketitle


\section{\label{sec:intro}Introduction}

Ever since the discovery of laser-induced ultrafast spin dynamics by Beaurepaire et al.~\cite{Beaurepaire1996} and subsequent realization of deterministic all-optical magnetic switching (AOS)~\cite{Stanciu2007}, the field of optically driven magnetization dynamics on the femtosecond time scale has become of great interest for two main reasons. 
First, understanding the fundamental mechanisms of non-equilibrium, ultrafast spin dynamics, and second, the potential application in the next generation of information technology with a vision for both faster and more energy efficient data storage devices. 
The understanding of temporal control of magnetic order has progressed rapidly~\cite{Koopmans2010}, with investigations in laser driven superdiffusive spin transport~\cite{Battiato2010, Rudolf2012, Shokeen2017} and its role in terahertz emission \cite{Kampfrath2013}, as well as very recently, the discovery of optically induced intersite spin transfer (OISTR)~\cite{Dewhurst2018,Siegrist2019, Hofherr2020, Willems2020}, which promises routes to tailor ultrafast magnetization dynamics via density of states engineering.
AOS, as revealed by experimental and theoretical studies, relies on ultrafast heating of the electronic system in the magnetic material, which directly modulates the inter-atomic exchange interaction and facilitates the magnetic moment reversal via exchange scattering \cite{Radu2011,Kirilyuk2013, Atxitia2013,Davies2020}. 
Importantly, AOS exhibits a strongly non-linear dependence on the excitation fluence and is characterized by a well-defined, narrow fluence window for which a deterministic reversal of the magnetization direction is observed. 
\par
\begin{figure*}[t]
  \centering
    \includegraphics{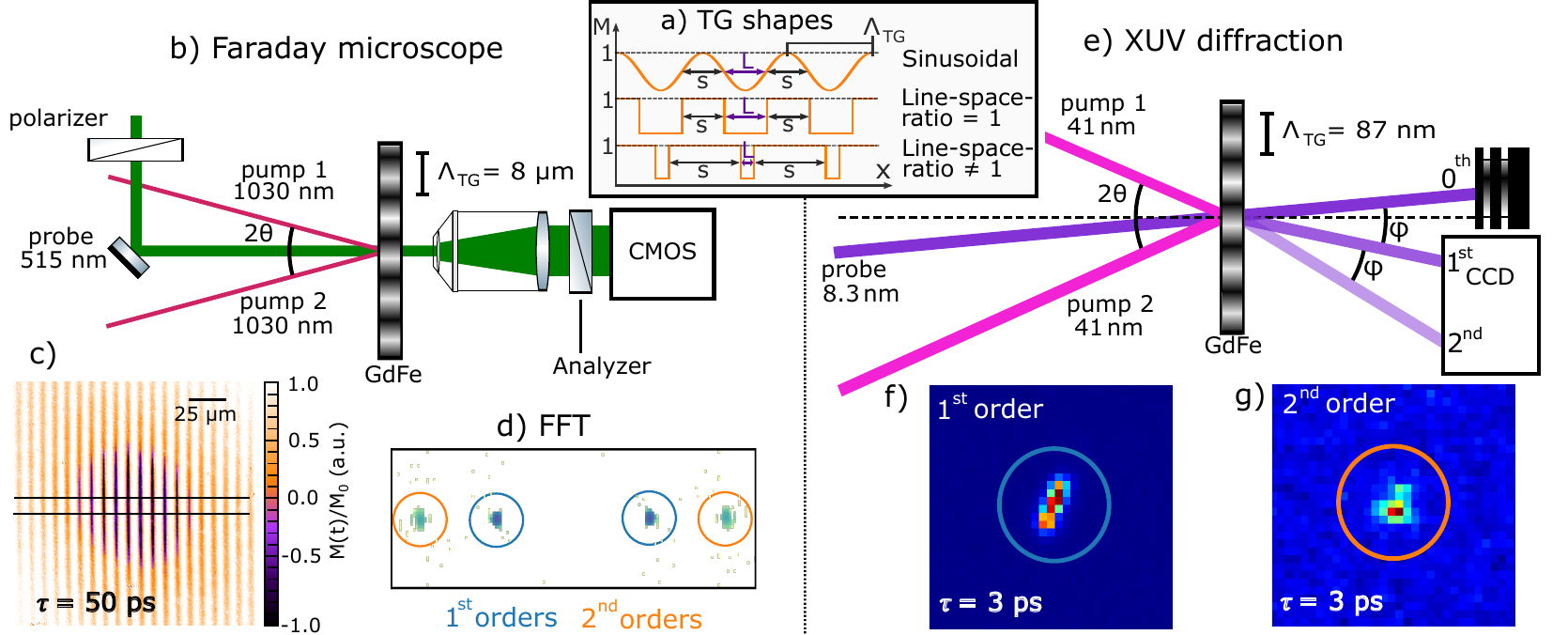}
  \caption{Experimental setups: a) Schematic examples of transient magnetic grating shapes with different line (l) and space (s) widths: sinusoidal, exhibiting only the first Fourier component; equal line and space width (line-space-ratio\,=\,1)  with suppression of even Fourier components; and unequal line and space widths (line-space-ratio\,$\neq$\,1), exhibiting the first, second and higher Fourier components. b) Wide-field magneto-optical microscope. Here, a TMG  with $\Lambda_\mathrm{TMG}=7.8$\,\textmu m is induced by interference of two femtosecond laser pulses in the near infrared spectral range. The evolving magnetization pattern is probed via the optical Faraday effect. c) Real-space image depicting the TMG at a delay of $\tau=50$\,ps. In the circular center, the excitation fluence is sufficient to induce AOS. d) Background-corrected Fourier transform showing the first and second Fourier coefficients of the magnetic TMG from panel c). e) Diffraction experiment at the free electron laser (FEL) facilities FERMI in Trieste, Italy. The interference of two XUV pump pulses generates a TMG with a periodicity of $\Lambda_\mathrm{TMG}=87$\,nm. The first f) and second g) order diffraction of a time-delayed, resonant XUV probe pulse at 8.3\,nm (150\,eV) are simultaneously recorded by a CCD camera.}
  \label{fig:fig1}
\end{figure*}
To realize all-optical-based magnetic reversal in technological applications, however, the understanding of light-induced spatial control of magnetization needs to be expanded into the nanometer range. 
One way to overcome the spatial limitations imposed by the wavelength of visible light when approaching the nanometer length scale, is to use nanostructures to confine the optical excitation.  
Examples include plasmonic enhancement of the optical driving field via metallic nanorods \cite{Xu2015,Choi2018} or gratings \cite{VonKorffSchmising2015}, revealing a significant increase in average demagnetization amplitudes in all-optical Kerr studies. 
Retrieving \textit{real-space} information on a nanometer spatial scale after localized optical excitation has remained challenging, though and only a limited number of experiments based on X-ray imaging techniques have been reported in literature \cite{VonKorffSchmising2014,Liu2015,LeGuyader2015,Zayko2021}.
On the other hand, \textit{reciprocal-space} information extracted from small angle X-ray experiments was, for example, able to successfully demonstrate the role of nanoscale inhomogeneities for AOS \cite{Graves2013}, underlining the importance of near-field enhancement in granular FePt layers in the field of heat-assisted magnetic recording \cite{Granitzka2017} and allowed investigations into nanoscale spin transport in a nano-patterned CoPd multilayer~\cite{Weder2020}.
\par
Another approach to gain access to ultrafast processes on a nanometer length scale is to reduce the excitation wavelength to the extreme ultraviolet (XUV) spectral range and induce a spatially periodic excitation via interference of two FEL pulses.
This technique has been pioneered at the EIS-Timer beamline of the free-electron laser (FEL) FERMI in Trieste, Italy~\cite{Bencivenga2019} and has been very recently employed for a first investigations of magnetization dynamics~\cite{Ksenzov2021}. 

In this study, we expand XUV transient grating spectroscopy by analyzing the combined ultrafast evolution of the first and second order diffraction as well as calibrate the \textit{reciprocal-space} observables by complementary measurements in \textit{real-space}. 
We excite a 20\,nm thick ferrimagnetic GdFe alloy sample by overlapping two XUV beams and generate a transient magnetic grating (TMG) with a periodicity of $\Lambda_\mathrm{TMG}=87$\,nm.
The spatial evolution of the magnetization grating is probed by diffracting a time-delayed, third XUV pulse tuned to the Gd \textit{N}-edge at a wavelength of 8.3\,nm (150\,eV).
Due to the non-linear fluence dependence of AOS, we expect a modified line-space-ratio of the evolving magnetic grating from the initial sinusoidal excitation pattern.
This information is directly encoded in the diffraction pattern, such that the simultaneously measured first and second diffraction orders allows us to demonstrate the emergence of AOS. 
To corroborate these results, we perform supporting all-optical real-space measurements on the same sample using a wide-field magneto-optical microscope in the visible spectral range~\cite{steinbach2021}, where we directly image how the periodic excitation evolves into a pattern with reversed magnetization. 
A Fourier analysis yields the corresponding signal in reciprocal-space and serves as a reference to quantify the XUV diffraction experiments.
In particular, we show that the ratio between the second and the first order diffraction intensities is a universal quantity sensitive to the shape of a grating and exhibits identical absolute numbers across both experiments and is therefore a precise observable with which we demonstrate AOS with a nanometer spatial and femtosecond temporal resolution.



\begin{figure}[!tbp]
  \centering
    \includegraphics{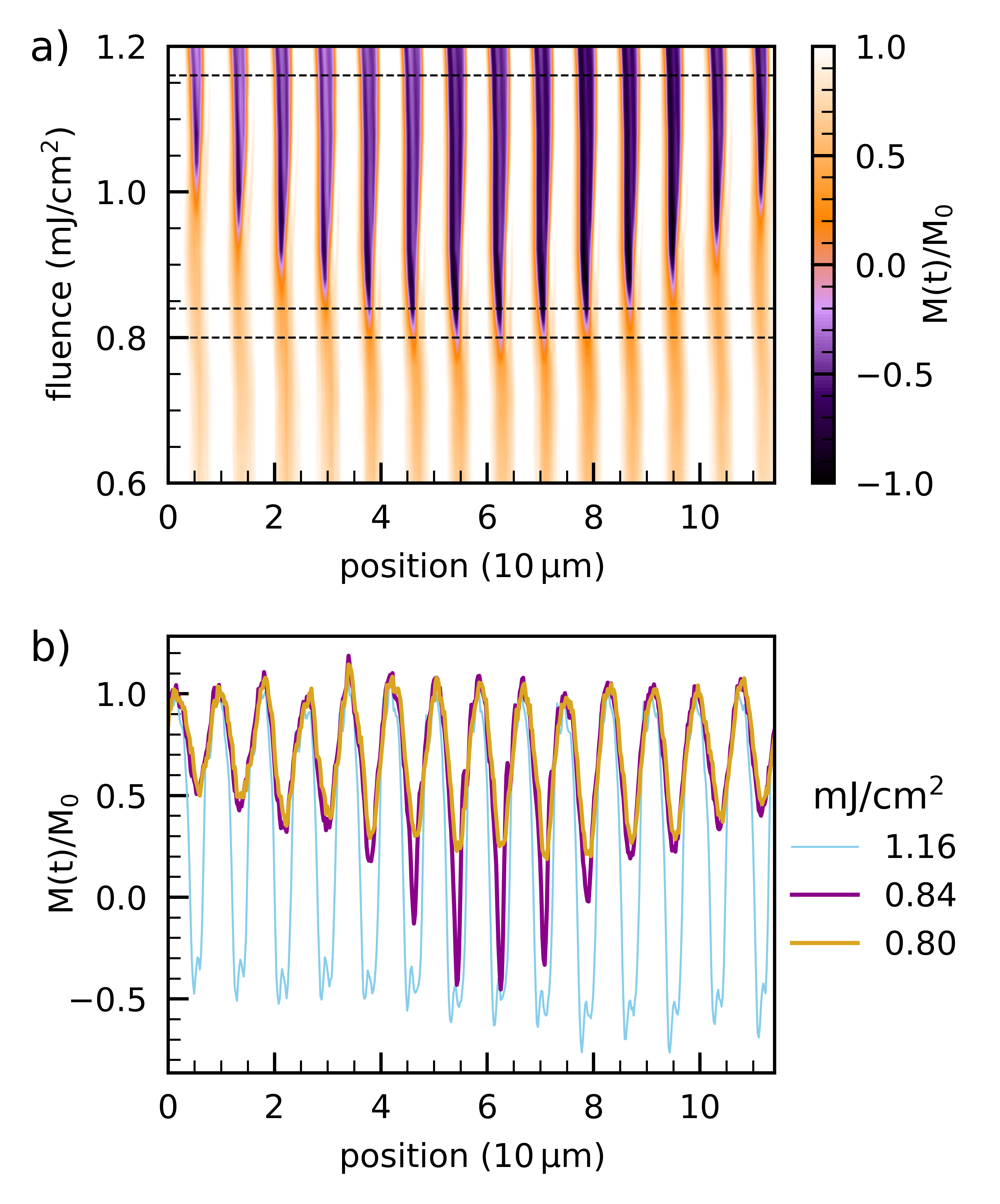}
  \caption{a) Lineouts of cropped real-space images as a function of pumping fluence (each of the two pump beams) at a delay of 50\,ps. b) Three lineouts showing the 1-D TMGs for three excitation fluences, as indicated by the horizontally dashed lines in a).
  }
  \label{fig:fig2}
\end{figure}


\section{\label{sec:results}Results}

The inset Fig.~\ref{fig:fig1}\,a) depicts the concept of our experiment: TMGs are characterized by demagnetized lines (l) and fully magnetized spaces (s), whose relative widths determine the observable diffraction orders, or Fourier components.
Any grating with identical widths of lines and spaces or a line-space-ratio\,=\,1 only exhibits odd Fourier components (diffraction orders) and suppressed even components.
Thus, the observation of an emerging second order diffraction corresponds to a deviation from the initial sinusoidal excitation pattern with a line-space-ratio\,$\neq$,1.
Hence, we can deduce information on the evolving grating shape by evaluating the time-dependent ratio between the second and first order diffraction intensities R$_{21}$.  
The experimental setups are schematically depicted in Fig.~\ref{fig:fig1}, where b) shows the Faraday microscopy setup using optical pump and probe pulses and e) shows the measurement geometry at the FERMI facility in Trieste, Italy using the EIS-TIMER beamline \cite{Bencivenga2019}, providing laser pulses in the XUV spectral range. 
In both experiments, two pump pulses are crossed under a total crossing angle $2\theta$ on the sample exciting a periodic interference pattern in shape of a sinusoidal grating with a line-space-ratio of 1 (equal line and space widths) and a period $\Lambda_\mathrm{TGM}=\lambda_\mathrm{pump}/2\sin(\theta)$ where $\lambda_\mathrm{pump}$ is the pump wavelength~\cite{Mishina1974}.

The investigated sample consists of an out-of-plane magnetized, 20\,nm thick Gd$_{24}$Fe$_{76}$ alloy thin film with 3\,nm thick Ta seed and cap layers, magnetron sputtered on a 30\,nm thick Si$_{3}$Ni$_{4}$ membrane and glass substrate within the same growth cycle.
The film is magnetized to saturation with an external magnetic field of 60\,mT aligned perpendicularly to the sample plane.
\par

The Faraday microscope provides femtosecond temporal and sub-micrometer spatial resolution~\cite{steinbach2021}. 
We overlap two pump beams ($\lambda = 1030$\,nm) under $2\mathit{\theta}=7.6^{\circ}$, exciting a TMG with a period of 7.8\,$\mu$m and directly record the real-space image with a third, time delayed femtosecond probe pulse ($\lambda = 515$\,nm) via the Faraday effect. 
By determining the absolute values of the magneto-optical polarization rotation, we calibrate the detected light intensity to the magnetization of the sample~\cite{steinbach2021}. 
An example of a real-space image of the transient magnetization pattern is shown in Fig.~\ref{fig:fig1}\,c) for a time delay of 50\,ps and pump fluence\footnote{Note, that all fluence values for the microscopy experiments are given for each pump pulse separately and refer to the pulse energy normalized by the full width at half maximum beamsize of the pump pulses.} of $0.92$\,mJ/cm$^2$. 

We observe a TMG with a well-defined circular area in which the GdFe sample reverses its magnetization, a clear indication of the strongly non-linear fluence-magnetization relationship with a sharp threshold fluence for which AOS can be induced. 
The background-corrected Fourier transform of the measured TMG images yields the corresponding transient reciprocal-space information.
An example is shown in Fig~\ref{fig:fig1} d) for the delay $\mathrm{\tau}=50$\,ps, from which we calculate the ratio between the integrated second and first order intensity, R$_{21}$. 
\par
In Fig.~\ref{fig:fig2}\,a), we show magnetization maps generated by averaging the spatially centrosymmeteric real-space image along the vertical dimension between the two horizontal lines as shown in Fig.~\ref{fig:fig1}\,c) for measurements with varied excitation fluence between 0.6\,mJ/cm$^2$ and 1.2\,mJ/cm$^2$ at a fixed delay of 50\,ps. 
The changing shape of the grating is visualized as lineouts at three different fluences in Fig.~\ref{fig:fig2}\,b): for 0.80\,mJ/cm$^2$, we achieve a demagnetization of about 70\% exhibiting a sinusoidal spatial profile with equal line and space widths following the excitation pattern.  
A small further increase of the fluence to 0.84\,mJ/cm$^2$ induces AOS and strongly affects the symmetry of the grating: the narrow switched regions now lead to a line-space-ratio \,$\neq$~1.
\begin{figure}[!tbp]
  \centering
    \includegraphics{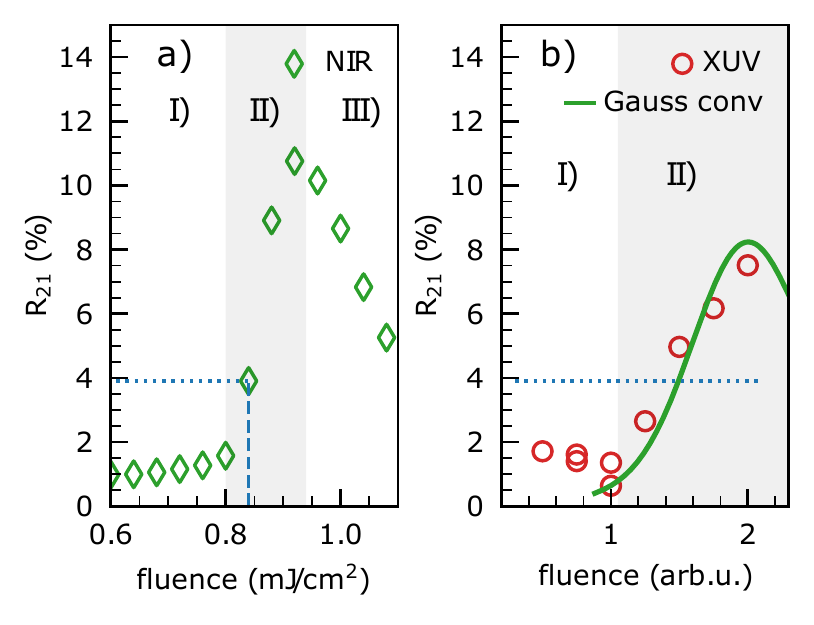}
  \caption{Ratio R$_{21}$ between the 2$^{nd}$ and 1$^{st}$ order as a function of the pump fluence at a delay of 50\,ps of a) Fourier transform of the real-space images excited with NIR light in green diamonds and b) the XUV diffraction experiment in red circles; the green line shows a linear interpolation of the NIR data convoluted with a Gaussian function with $\sigma=0.3$.}
  \label{fig:fig3}
\end{figure}
An even stronger excitation of 1.16\,mJ/cm$^2$ results in a square shape of the grating, however, again yielding a line-width-ratio of approximately 1.
In Fig.~\ref{fig:fig3}\,a), we show the corresponding values of R$_{21}$.
Three distinct regimes of R$_{21}$ are visible.
In the interval (I) below 0.80\,mJ/cm$^2$, where the sample is demagnetizing, but not yet switched, R$_{21}$ remains constant at about 1-2\%.
From 0.84 to 0.94\,mJ/cm$^2$ (II), where R$_{21}$ rises to 11.0\%, we start to switch in the center of the excitation and finally, from 0.98 to 1.2\,mJ/cm$^2$ (III), where R$_{21}$ drops back down to 1.1\%, the switched lines become wider such that the grating profile once again approaches a line-width-ratio of 1.

\begin{figure}[!tbp]
  \centering
    \includegraphics{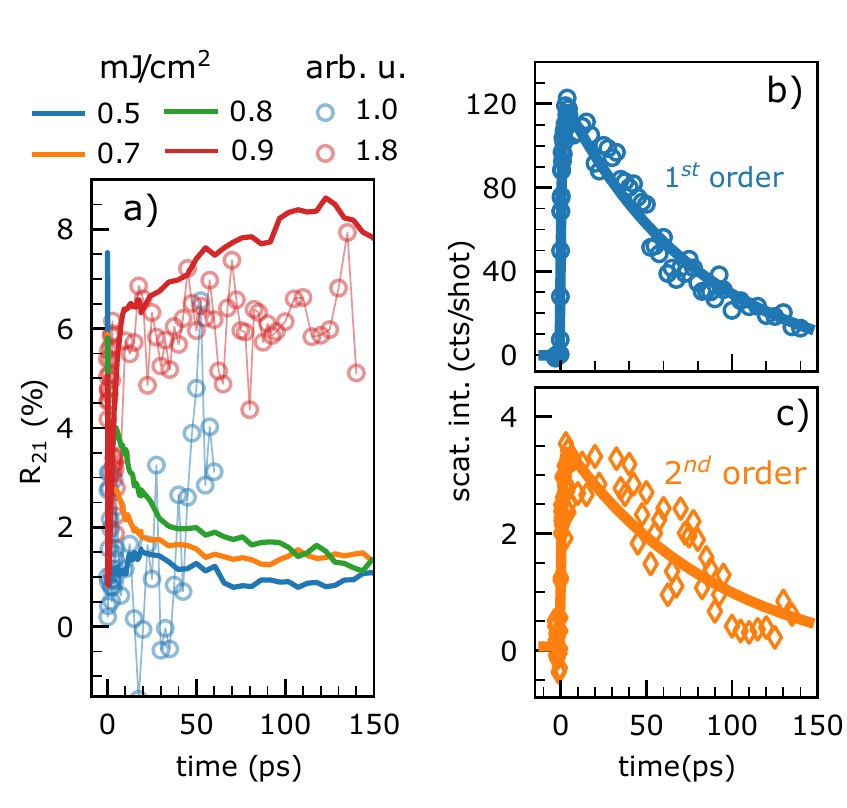}
  \caption{Diffracted XUV intensity off the TMG in counts per shot as a function of the time delay between the pump and probe pulses for a) the 1$^{st}$ (blue) and b) the 2$^{nd}$ order diffraction intensity (orange), fitted with double exponential functions. c) Ratio R$_{21}=I_{2^\mathrm{nd}} /I_{1^\mathrm{st}}$ between the 2$^\mathrm{nd}$ and 1$^\mathrm{st}$ order diffraction intensity as a function of delay. The open circles represent XUV diffraction data for a strong (red circles) and a weaker excitation density (blue circles). The solid lines depict the magneto-optical microscope data, where we observe a qualitative change in the evolution between 0.8 (solid green) and 0.9\,mJ/cm$^2$ (solid red). }
  \label{fig:fig4}
\end{figure}

In Fig.~\ref{fig:fig4}\,a), we show the temporal evolution of R$_{21}$ extracted from the Faraday measurements in solid lines for different pump fluences ranging from 0.5 to 0.9\,mJ/cm$^2$.
We observe that for excitations leading to demagnetization only (up to \SI{0.80}{\fluence}), R$_{21}$ initially rises and then rapidly relaxes to small values and remains constant on longer time scales.
However, when AOS occurs with an excitation fluence of 0.9\,mJ/cm$^2$, the temporal evolution of R$_{21}$ changes drastically: now the ratio increases within picoseconds and reaches up to 8\%  on longer time scales.


In the corresponding diffraction experiment at FERMI, the pump beams with wavelengths $\lambda_\mathrm{pump}=41.4$\,nm (30\,eV) are crossed under $2\mathit{\theta}=27.6^{\circ}$ resulting in an excitation pattern with a grating period $\Lambda_\mathrm{TGM}=87$\,nm.
This periodic excitation leads to a transient change of the magnetization and a corresponding modulation of the magnetic circular dichroism (MCD)~\cite{Ksenzov2021}, allowing to probe the magnetization by diffraction of a third, time-delayed linearly polarized XUV probe beam tuned to the Gd \textit{N}$_{4,5}$-edge at a wavelength of 8.3\,nm. 
The first (Fig.~\ref{fig:fig1}\,f)) and second (Fig.~\ref{fig:fig1}\,g)) diffraction orders are simultaneously recorded by a charge coupled device (CCD) placed at a distance of 150\,mm behind the sample.
\par

\begin{figure}[!tbp]
  \centering
    \includegraphics{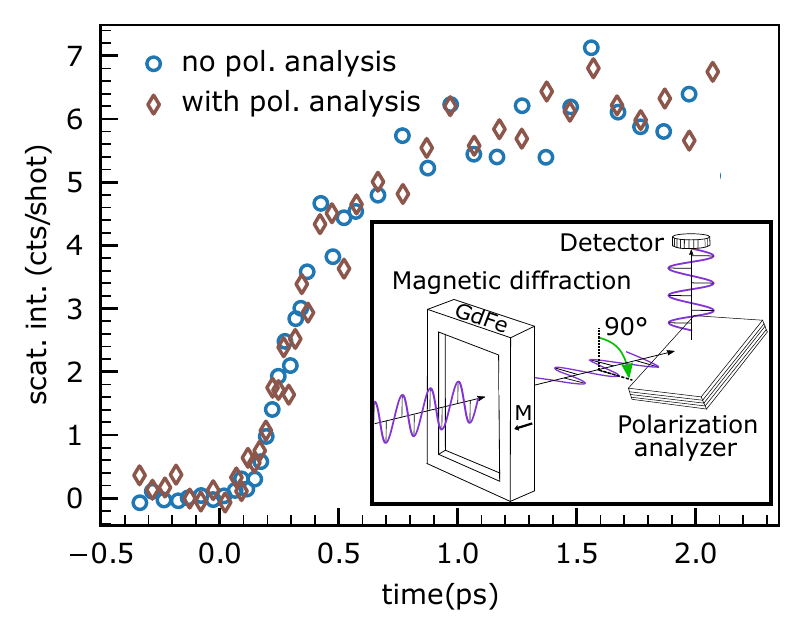}
  \caption{Comparison of the early temporal evolution of the resonantly diffracted XUV probe pulse off the TMG at the Gd \textit{N}$_5$ edge at \SI{150}{\electronvolt} with no polarization analysis (red circles) and with polarization analysis (green diamonds), scaled by a factor of 12.5. The agreement indicates that the recorded signal without polarization is predominately of magnetic origin.}
  \label{fig:fig5}
\end{figure}

Because magnetic scattering leads to a 90 deg rotation of the polarization \cite{Kortright2013}, we first confirm the magnetic origin of our signal with a polarization analysis using a removable multilayer mirror set at the Brewster angle of $\approx45^{\circ}$.
The inset in Fig.~\ref{fig:fig5} schematically depicts the geometry: the diffracted probe beam is reflected vertically upwards from the analyzer into a second CCD camera. 
We show the early time evolution of the diffracted intensity off the TMG in counts per FEL shot without polarization analysis (blue circles) and with polarization analysis (brown diamonds). 
For each image, i.e. delay point, we accumulate the scattering intensity for up to 4000 shots.
The comparison demonstrates that our diffracted signals are indeed of magnetic origin. 
To compensate for reflection losses the data recorded via the analyzer is multiplied with a factor of 12.5.
We note that potential contributions to the diffraction intensity due to changes of the electro-optical constants are expected to be small for the probed 4\textit{f} states of Gd
\footnote{Due to geometrical constraints of the setup, this experiment was performed with pumping wavelengths of 20\,nm resulting in transient gratings with $\Lambda_\mathrm{TGM}=54$\,nm.}.
\par

\par
In Fig.~\ref{fig:fig4}, we show the evolution of b) the first and c) the second order diffraction signals on a longer time scale up to 150\,ps. 
We find comparable decay times of $\tau_{\searrow1st}=(81\pm7)$\,ps and $\tau_{\searrow2nd}=(90\pm24)$\,ps.
Here, the pump fluence was set to 1.8\,arb.u. which was close to the highest possible fluence without damaging the sample.
\par
\par
While the TMG decays within 150\,ps in the majority of the probed area due to magnetization recovery (cf. first and second orders Fig.~\ref{fig:fig4}\,b)\,\&\,c)), the corresponding ratio R$_{21}$ (Fig.~\ref{fig:fig4}\,a) red circles) remains constant at about 6\% in the same time interval.
This reveals a more stable magnetic structure and a deviation of the line-space-ratio from 1; the first indication for AOS.
On the other hand, when the pump fluence is too low to induce AOS (Fig.~\ref{fig:fig4}\,a) blue circles), R$_{21}$ remains constant at low values of approximately 1\%, independent of the time delay.
\footnote{Note, that the larger fluctuations of R$_{21}$ are due to the very small signals in the second order diffraction intensity and the small first order photon counts for later times.}
These observations confirm, that the TMG amplitude scales linearly for low excitation fluences, where equally sized magnetized spaces and demagnetized lines suppress the even diffraction orders.
As soon as the excitation fluence exceeds the switching threshold, we enter the non-linear regime between excitation and magnetization amplitude, causing an increase of R$_{21}$, due to a reduction of the line-space-ratio of the TMG.

In Fig.~\ref{fig:fig3}\,b), we show the fluence dependency of R$_{21}$ at a fixed delay of 50\,ps in the XUV scattering experiment, characterized by two distinct regimes: (I) constant R$_{21}=1.7\%$ up to fluence of 1.0\,arb.u. followed by (II) a rise up to R$_{21}=7.9\%$. 
For even higher excitation fluences we started to observe a permanent grating structure indicative of sample damage.
The increased values of R$_{21}$ in interval (II) provide a second piece of evidence for AOS in the transient grating experiment.
Closer inspection of Fig.~\ref{fig:fig3} reveals that in the XUV scattering experiment the increase of R$_{21}$ extends over a broader range of fluences compared to the measurements in IR. 
However, this is expected taking into account the pump pulse fluctuations of the FEL source. 
Such pump pulse fluctuations imply that each measurement at a particular fluence setting is actually composed of diffractions off different magnetic gratings with and without AOS. 
Indeed, a convolution of the values of $R_{21}$ retrieved in the optical measurements with a Gaussian function with $\sigma=0.3$, yields the solid green line in Fig.~\ref{fig:fig3}\,b) and overlaps with the XUV scattering data. 
The assumed value of the standard deviation is in good agreement with the average fluctuations of the FEL pulses. 

\par

To further justify the direct comparison of R$_{21}$ extracted from the reciprocal and real-space data, we briefly discuss the two main differences of the experiments, namely the distinct TMG periodicity, $\Lambda_\mathrm{TMG}$, of 87\,nm vs. 7.8$\mu$m and the distinct excitation photon energies of 30.0\,eV vs. 1.2\,eV.
Regarding the different length scale of the grating periodicities, we note that AOS is triggered by a transient, hot electron distribution, such that the lateral spatial limit for AOS is expected to be determined by the mean free path of photo-excited, spin-polarized electrons. 
As their values are limited by the large DOS at the Fermi energy of GdFe and are typically at least an order of magnitude lower than the XUV induced grating size of $\Lambda_\mathrm{TMG} = 87$\,nm \cite{Zhukov2006,Chen2019, Weder2020}, we do not expect that ultrafast electron transport significantly alters the shape of the TMGs. 
On the other hand, the observed quantitative agreement between micro- and nanoscale gratings, suggests that we have not yet reached the fundamental spatial limits for AOS. 
Secondly, while high intensity XUV photon-matter-interaction is a novel and not well explored area of research, it is generally assumed that electron-electron scattering very rapidly thermalizes the initially strongly non-equilibrium electron distributions, leading to comparable demagnetization and AOS processes as observed in the optical regime \cite{Higley2019, Liu21}. 

Finally, we would like to point out that significantly smaller optically reversed domains then given by $\Lambda_\mathrm{TMG}$ can be achieved by taking advantage of the well-defined fluence window for which AOS is observed. 
For a sinusoidal excitation, this can be realized either by tuning the maximal fluence just above the switching threshold or by increasing the maximal fluence beyond the switching window leading to more complicated double line substructures. This can be easily appreciated by inspection of Fig.~\ref{fig:fig2} for the two lineouts at a fluence of 0.84\,mJ/cm$^2$ and 1.16\,mJ/cm$^2$. 
Importantly, such substructures observed for higher fluences can be again identified in diffraction experiments via a distinct evolution of R$_{21}$ as a function of fluence. 
As confirmed by simulations, the presence or absence of substructures leads to a decrease or unchanged values of R$_{21}$ in interval (III) (Fig.~\ref{fig:fig3}\,b)), respectively.

In summary, we have, for the first time, demonstrated the emergence of all-optical switching on the nanometer length scale in a GdFe sample by overlapping two XUV pump pulses in a transient grating spectroscopy experiment with femtosecond time resolution.
We reveal the non-linear demagnetization response as a function of excitation fluence, a fingerprint of AOS, via the ratio between the second and first order diffraction intensities, R$_{21}$, which is particularly sensitive to changing line-space ratios of the TMG.
Moreover, we directly image the evolution of AOS in TMG experiments on the micrometer length scale via transient Faraday microscopy and extract corresponding diffraction ratios R$_{21}$ via Fourier analysis. 
Quantitative agreement between the optical imaging and XUV diffraction experiments allow us to connect the nanoscale reciprocal and micrometer real-space information, further supporting our finding of the ultrafast emergence of AOS on the nanometer length scale.

\begin{acknowledgments}
C.v.K.S. and S.E. would like to thank DFG for funding through TRR227 project A02. We acknowledge Stefano Bonetti for providing the electro-magnet used during the experiment.

\end{acknowledgments}

\appendix

\bibliography{biblio}

\end{document}


\preprint{AIP/123-QED}

\title[]{Supplemental: All-optical switching on the nanometer scale excited and probed with femtosecond extreme ultraviolet pulses}

\author{Kelvin Yao}
 \affiliation{Max-Born-Institut für Nichtlineare Optik und Kurzzeitspektroskopie, Max-Born-Straße 2A, 12489 Berlin, Germany}
\author{B. Author}%
 \email{Second.Author@institution.edu.}
\affiliation{ 
Authors' institution and/or address
}%

\author{C. Author}
 \homepage{http://www.Second.institution.edu/~Charlie.Author.}
\affiliation{%
Second institution and/or address
}%

\date{\today}

\begin{abstract}
Ultrafast control of magnetization on the nanometer length scale, in particular all-optical switching (AOS), is key to putting ultrafast magnetism on the path 
towards future technological application in data storage technology. 
However, magnetization manipulation with light on this length scale is challenging due to the wavelength limitations of optical light.
Here, we show AOS on the nanometer scale by interfering two extreme ultraviolet (XUV) pulses from a free electron laser (FEL) at the FERMI facilities, exciting transient magnetization gratings (TG) with periods of 86.8\,nm and resonantly scatter from the TG at the Gd \textit{N}-edge, probing the magnetic excitation in a GdFe alloy.
By examining the simultaneously recorded first and second order scattering and by performing reference real-space measurements on a wide-field magneto-optical microscope, we can conclusively demonstrate AOS on the nanometer scale.

\end{abstract}

\maketitle

\section{\label{sec:supplemental}Supplemental}

\begin{figure}
  \centering
    \includegraphics{./images/apdx1.pdf}
  \caption{Simulation of the excitation interference pattern on the sample with $\Lambda_{TG}=86.6$\,nm analogous to the FERMI scattering experiment (orange axis label) and $\Lambda_{TG}=7.8$\,$\mu$m analogous to the Faraday microscopy experiment (blue axis label), both with a peak fluence of 5.4\,$\frac{\text{mJ}}{\text{cm}^{2}}$. This excitation pattern is used to simulate a magnetization response using a fluence map \cite{steinbach2021widefield} which is then Fourier transformed for the first and second orders of the magnetization grating.}
  \label{app:1}
\end{figure}

\begin{figure}[!tbp]
  \centering
    \includegraphics{./images/apdx2.pdf}
  \caption{Difference between the second and first order ratios r$_{21}$ from \ref{app:1} of the simulated FERMI experiment and the micrscopy experiment with r$_{21difference}=r_{21FERMI}-r_{21Microscope}<\pm0.1\%$.}
  \label{app:2}
\end{figure}

\begin{figure*}[!tbp]
  \centering
    \includegraphics{./images/apdx3.pdf}
  \caption{First (a)) and second order (b)) temporal progressions extracted from the real-space Faraday microscopy images via Fourier transform for different pump fluences (refering to each pump arm), corresponding to r$_{21}$ in Fig~\ref{fig:fig3} b) in solid lines.}
  \label{app:3}
\end{figure*}

\begin{figure}[!tbp]
  \centering
    \includegraphics{./images/apdx4.pdf}
  \caption{Complete fluence dependent r$_{21}$ with four intervals of behaviour up until sample breakage.}
  \label{app:4}
\end{figure}

\begin{figure}[!tbp]
  \centering
    \includegraphics{./images/apdx5.pdf}
  \caption{Fluence dependent r$_{21}$ response for a non-switching sample until sample breakage showing only two intervals of behaviour at a delay of 50\,ps [[sample name]].}
  \label{app:5}
\end{figure}

In order to confirm, that the observations in reciprocal space of the corresponding real-space images from the Faraday microscope give transferable knowledge about the magnetization state in the scattering experiment, we perform simulations using an excitation pattern shown in Supl-Fig.~\ref{app:1}, where the blue axes reflect the excitation pattern for the Faraday microscopy ($\Lambda_{TG}=7.8$\,$\mu$m) and orange axes the pattern for the scattering experiment at FERMI ($\Lambda_{TG}=86.6$\,nm). 
The excitation fluence in $\frac{\text{mJ}}{\text{cm}^{2}}$ is then mapped to a corresponding magnetization state at a select delay using a fluence map~\cite{steinbach2021widefield}, which yields a simulated magnetization TG, ready for Fourier analysis.
In Supl-Fig.~\ref{app:2}, we show the difference between r$_{21}$ of the simulated magnetization dynamics of the TGs on a nanometer and on the micrometer scale and find a absolute difference of below 0.1\%, which we attribute to numerical errors of the simulation, demonstrating that r$_{21}$ is a robust observable across experiments with differen TG periods. 

In Supl-Fig.~\ref{app:3} a) and b), we show the temporal progression of the first and second TG orders, respectively for different pump fluences, obtained by Fourier transformation of the real-space images recorded with the Faraday microscope.
The ratio r$_{21}$ between the second and first order is shown in Fig~\ref{fig:fig3} b) in solid lines.
With this figure, we would like to demonstrate that it is difficult to distinguish between pure demagnetization and partial AOS, when only the first scattering order is examined.
Even though AOS already occurs with a pump fluence of 0.9\,$\frac{\text{mJ}}{\text{cm}^{2}}$ (red solid line) in the center of the excitation, where the excitation is the strongest, there is no strong indication of such in the first order compared to the temporal progression of the non-switching 0.7 and 0.8\,$\frac{\text{mJ}}{\text{cm}^{2}}$ excitation (orange and green solid lines respectively). 
Only when the second order scattering is included, do we see a clear lingering scattering intensity for more than more than 100\,ps due longer time stability of switched areas compared to demagnetized areas(Fig.~\ref{app:1} b).
As shown in Fig~\ref{fig:fig3} b) and discussed in~\ref{sec:results}, the ratio r$_{21}$ yields even more obvious indications for AOS, thereby demonstrating its effectiveness as an observable.
In case of AOS across the whole probing area due to high pump fluence, the first order scattering is already enough to expose AOS with an almost non-decaying scattering signal within 100\,ps, as shown in Fig.~\ref{app:1} a) with 1.0 and 1.2\,$\frac{\text{mJ}}{\text{cm}^{2}}$ pump fluence (purple and brown solid lines, respectively).
However, the second order once again reveals further information: While the maximum scattering signal in the first order increases with increased pump fluence, reflecting the magnetic contrast of the TG, the second order scattering intensity diminishes at the highest pump fluence. 
As already discussed in section~\ref{sec:results}, this is due to structural change within the magnetic TG. 
[[Substructure]]

In Fig~\ref{fig:fig4} b) and section~\ref{sec:results}, we halve already shown and discussed r$_{21}$ in dependence of the excitation fluence in intervals 1), II) and III).
In Fig~\ref{app:4}, we show the complete fluence dependence, where we increased the pump fluence until sample breakage. 
This reveals another interval, which is labeled iV) in Fig~\ref{app:4}, where r$_{21}$ increases with higher fluence due to the indefinite broadening of the lines, both switched and demagnetized. 
Similarly, in samples which do not exhibit AOS, the magnetization reaches a saturated demagnetization level at a sufficient excitation fluence, which causes a continuous broadening of the lines and therefore a continuous increase of the second order relative to the first, shown in interval II) in Fig~\ref{app:5}.

\begin{acknowledgments}

\end{acknowledgments}

\appendix

\bibliography{biblio}